\documentclass[12pt]{iopart}

\usepackage{graphicx}
\usepackage{amssymb}

\begin{document}

\title[]{Neutrino oscillations as a novel probe for a minimal length}
\author{Martin Sprenger}
\ead{sprenger@fias.uni-frankfurt.de}
\author{Piero Nicolini}
\ead{nicolini@th.physik.uni-frankfurt.de}
\author{Marcus Bleicher}
\ead{bleicher@th.physik.uni-frankfurt.de}
\address{Frankfurt Institute for Advanced Studies (FIAS) \& Institut f\"{u}r Theoretische Physik, Johann Wolfgang Goethe-Universit\"{a}t,
60438 Frankfurt am Main, Germany}

\begin{abstract}
We suggest that the presence of a quantum gravity induced minimal length can be explored using neutrino oscillation probabilities.
Neutrinos seem ideally suited for this investigation because they can propagate freely over large distances and can therefore pile up minimal length effects beyond detectable thresholds.
We determine the modified survival probability in a scenario with a minimal length and find deviations from the classical behaviour for high energies.
We find that for the currently available experimental statistics the deviations from the standard oscillations do only allow for a bound of $\ell^{-1}\gtrsim 10~\mathrm{GeV}$ from MINOS data.
On the other hand, oscillations of high-energy neutrinos emitted by galactic and extragalactic sources are strongly suppressed, leading to a possible observation of quantum gravity effects at neutrino telescopes such as IceCube and ANTARES.
\end{abstract}

\maketitle

\section{Introduction}
Quantum gravity has been drawing the attention of the scientific community for more than thirty years, but it is still far from being fully understood and explained.
While there is some progress on the theoretical side due to the presence of at least two viable candidate theories, string theory and loop quantum gravity, the absence of any experimental quantum gravity data up to now is still puzzling \cite{AmelinoCamelia:1999zc}.
The main difficulty in getting experimental evidence is connected to the extremely large value of the Planck energy scale $E_P\sim 10^{16}~\mathrm{TeV}$ at which quantum gravity signals are supposed to appear.
This is 15 orders of magnitude higher than the energy scale tested by the LHC.
However, the arguments leading to this estimation rely on the assumption that general relativity holds up to the Planck scale.
New physics might enter much earlier and make quantum gravity effects important at much lower energy scales.\\
In recent years a new field has emerged that does not aim at a fundamental description of quantum gravity but instead uses effective theories to study quantum gravity effects phenomenologically.
Typically, one starts from the conventional theories, general relativity or quantum field theory, and implements an effect that is motivated by a fundamental theory.
A nice overview over the field of quantum gravity phenomenology can be found in \cite{Hossenfelder2010zj}.\\
The effect we will concern ourselves with is the emergence of a minimal length.
This idea is dated back to early times of quantum gravity \cite{DW} and emerges in all formulations of quantum gravity (cf. \cite{ML_LQG}, \cite{ML_ST}, \cite{Szabo2010}).
%This last fact allows us to make predictions \textbf{that are not linked to any fundamental theory} by employing an effective theory instead of studying the effects in a specific fundamental approach.
Different approaches to implement a minimal length exist and give rise to a model of quantum spacetime.
For instance, generalised uncertainty principle (GUP) approaches \cite{Kempf:1994su}, \cite{Kempf_QFT} start from a modified commutator
\begin{equation}
\left[\hat{x},\hat{p}\right]=i\left(1+f(\vec{p}^{\, 2})\right)
\label{eq:gup_comm}
\end{equation}
and work in a momentum representation of the corresponding algebra.
In this paper, we will work with a non-commutative geometry (NCG) approach (for a review of other approaches see \cite{ReviewNCG}).
In \cite{NCQFT} an original formulation based on coordinate coherent state NCG has been successfully employed to improve classical curvature singularities in black hole spacetimes \cite{nonlocal,NCBHs} and provide a reliable description of the fractal structure of the universe at the fundamental scale \cite{fractal}.
Starting from a commutator 
\begin{equation}
\left[\hat{x}^\mu,\hat{x}^\nu \right]=i\theta^{\mu\nu}
\end{equation}
a representation of the corresponding algebra in terms of coherent position states is built.
In this representation, delta functions appearing in particle field theories are widened into Gaussian distributions.
Equivalently to starting from the coherent states, it has been shown that the non-commutative smearing effect can be represented by the action of a non-local operator $e^{\ell^2 \Delta_x}$ which spreads the point-like Dirac delta into a Gaussian distribution, as well, i.e.
\begin{equation}
e^{\ell^2 \Delta_x}\delta(\vec{x})=\rho_{\ell}(\vec{x})=\frac{1}{(4\pi\ell^2)^{d/2}} e^{-\vec{x}^{\,2}/4\ell},
\label{spreading}
\end{equation}
where $\Delta_x$ is the Laplacian operator on a $d$ dimensional Euclidean manifold and $\ell$ is the NCG induced minimal length.
In momentum space, this corresponds to modifying the integration measure as 
\begin{equation}
d^{3}p\rightarrow e^{-\ell^2 p^2} d^{3}p.
\label{nlv}
\end{equation}
The leading order corrections to any field equation in the presence of a non-commutative background can then be obtained by replacing the conventional point like source term (matter sector) with a Gaussian distribution, while keeping differential operators (geometry sector) formally unchanged.
The main approach to non-commutative geometry via the Moyal $\star$-product suffers from the necessity to expand scattering amplitudes in the minimal length parameter, destroying the non-local character of the theory.
In contrast, our deformation in Eq.(\ref{nlv}) contains an infinite number of terms and is intrinsically non-local, leading to a UV finite and unitary field theory \cite{NCQFT,nonlocal}.\\
In the above formulation the minimal length is not set \textit{a priori}.
Typically, one expects $\ell$ to be of the order of the Planck length.
However, as mentioned above, quantum gravity effects might become stronger at much lower scales.
All we know from experiments is that $\ell\gtrsim 1~\mathrm{TeV}$ \cite{newton}.  
In this paper, we propose to study neutrino oscillations as an accurate phenomenon to test quantum gravity effects resulting from the minimal length.
A modified oscillation pattern could then be used to constrain the value of the minimal length.
Note again that this is a phenomenological study.
We do not provide a quantitative analysis but aim for a qualitative study of the minimal length effects and their possible observation.\\
High-energetic particles probe the structure of spacetime to microscopic scales and should be able to feel effects induced by a minimal length.
For this reason neutrino propagation has already been the subject of investigations for possible tests of quantum decoherence \cite{Lisi}, modified dispersion relations \cite{JChristian} and modified de Broglie relations \cite{GUPNeutrino}.
In addition, quantum gravity effects in neutrino physics have also been investigated by the IceCube collaboration \cite{Kelley:2009zza}.
In contrast to these studies, we study neutrino oscillations in a framework that has already been successfully employed to study minimal length effects in collider experiments \cite{signatures}, the Casimir effect \cite{casimir}, the Unruh effect \cite{ueff}, the magnetic moment of the muon \cite{g-2} and in black hole physics \cite{nonlocal,NCBHs}.\\
Even if the expected deviations are very small, neutrinos seem ideally suited for minimal length studies.
Neutrinos can propagate freely over astronomic distances without interacting with matter, giving rise to the hope that the propagation over long distances will pile up the minimal length effect beyond currently detectable thresholds.
As a simple analogy from mechanics, friction effects of a moving particle become significant for high velocities, but also for small velocities if the propagation distance is long enough.

\section{Neutrino oscillations}
The most common interpretation of neutrino oscillations is that neutrinos do not propagate in a flavour eigenstate but in a mass eigenstate.
Neutrinos are not massless, as assumed in the Standard Model, but have a non-vanishing mass.
The basis change from the flavour eigenbasis to the mass eigenbasis is described by a unitary $3\times 3~$-matrix $U$  (the Pontecorvo-Maki-Nakagawa-Sakata matrix) which is parametrized by three mixing angles $\theta_{12},~\theta_{13},~\theta_{23}$ and a CP-violating phase $\delta_{CP}$:
\begin{equation}
\left|\nu_\alpha\right\rangle=\sum\limits_{k=1}^{3}U_{\alpha k}^\ast\left|\nu_{k}\right\rangle,
\nonumber
\end{equation}
where Greek indices stand for flavour eigenstates $\left|\nu_\alpha\right\rangle$ while Latin indices stand for mass eigenstates $\left|\nu_k\right\rangle$.
Since the free Hamilton operator is diagonal in the mass eigenbasis we have
\begin{equation}
\left|\nu_k(t)\right\rangle=\exp(-iE_kt)\left|\nu_k\right\rangle,
\label{free}
\end{equation}
with $E_k$ being the energy of the neutrino and t the propagation time.
The oscillation propability for a flavour change from flavour $\alpha$ to flavour $\beta$ is therefore given by
\begin{eqnarray}
\nonumber P(\nu_\alpha\rightarrow\nu_\beta)&=&\left|\left\langle\nu_\beta|\nu_\alpha(t)\right\rangle\right|^2\\&=&\sum\limits^{3}_{k,j=1}U^\ast_{\alpha k}U_{\beta k}U_{\alpha j}U^\ast_{\beta j}e^{-i(E_k-E_j)t}.
\label{eq:osc_prob}
\end{eqnarray}
Here, we can approximate 
\begin{equation}
E_k-E_j=\sqrt{p^2+m_k^2}-\sqrt{p^2+m_j^2}\approx \frac{\Delta m_{kj}^2}{2E}.
\end{equation}
By setting $t\approx L$, where L is the propagation length, we get
\begin{equation}
P(\nu_\alpha\rightarrow\nu_\beta)=\sum\limits^{3}_{k,j=1}U^\ast_{\alpha k}U_{\beta k}U_{\alpha j}U^\ast_{\beta j}e^{-i\frac{\Delta m_{kj}^2}{2E}L}.
\label{eq:osc_app}
\end{equation}
From the quantity appearing in the exponential one defines the oscillation phase $\frac{\phi}{2\pi}=\frac{\Delta m^2}{2E}L=\frac{L}{L_o}$ with the oscillation length $L_o=\frac{2E}{\Delta m^2}$.
For large values of the phase, the oscillations become more and more rapid and the observable effect is washed out.
As a result, clear signals arise for small oscillation phases only.
However, for too small oscillation phases, oscillations do not occur.
Therefore, the oscillation length should be of the order of the propagation length.\\
In several experimental regimes, the oscillation reduces to a two-flavour problem (for details, see \cite{Winter2010}).
In this case the basis change from flavour eigenbasis to mass eigenbasis can be characterized by a single angle
\begin{equation}
U=\left(\begin{array}{cc}\cos\theta & \sin\theta\\-\sin\theta & \cos\theta\end{array}\right).
\label{eq:2f_matrix}
\end{equation}
The transition probability then simplifies to
\begin{equation}
P(\nu_\alpha\rightarrow\nu_\beta)=\sin^2(2\theta)\sin^2\left(\frac{\Delta m^2L}{4E}\right),
\label{eq:trans_prob}
\end{equation}
where $\Delta m^2$ is the difference of the squared masses of the two remaining mass eigenstates.
From that we find the survival probability $P(\nu_\alpha\rightarrow\nu_\alpha)=1-P(\nu_\alpha\rightarrow\nu_\beta)$.\\

\section{Neutrino oscillations with a minimal length}
\begin{figure}
\centering
\includegraphics[scale=0.9]{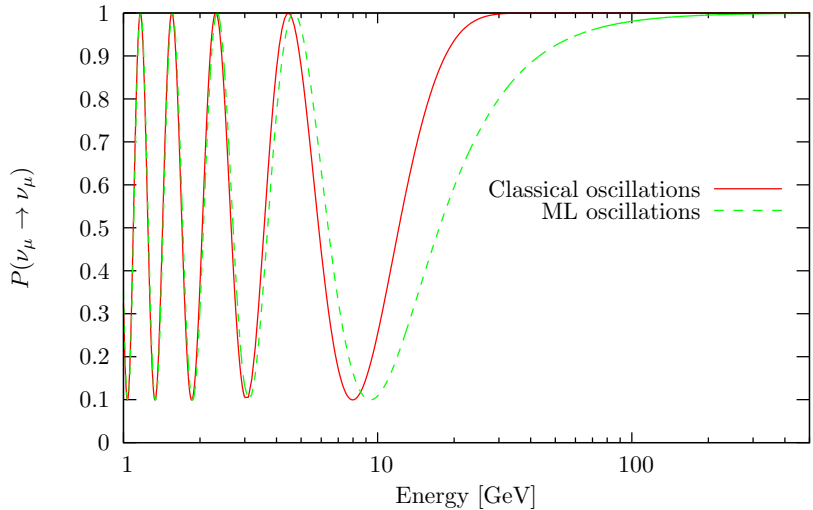}
\caption{Comparison of classical and minimal length oscillations for the two-flavour case with $\Delta m^2=2.3\cdot 10^{-3}~\mathrm{eV}^2$, $\sin^2(2\theta)=0.9$, baseline $L=5000~\mathrm{km}$ and $\ell^{-1}=10~\mathrm{GeV}$.}
\label{fig:comp_osci}
\end{figure}

Let us now implement the minimal length $\ell$.
In conventional quantum mechanics, wave number and momentum coincide, $k=p$.
From this, one finds that momentum eigenstates are plane waves in the position eigenbasis due to
\begin{equation}
\int dk'\,e^{ik'x}\delta(p(k)-p(k'))=\int dp'\,e^{ip'x}\delta(p-p')=e^{ipx},
\end{equation}
where for simplicity we are considering the one-dimensional case.
This situation changes in GUP models.
As explained in detail in \cite{hossenfelder}, the momentum remains unbounded in GUP theories.
The wavelength of a particle, however, is restricted to be larger than the minimal length $\lambda>\ell$, leading to a non-linear relation $k(p)$ between the wave vector and the momentum.
This is what also happens in the case of the coherent state approach to non-commutative geometry \cite{NCQFT}.
As mentioned above, all modifications due to the non-commutativity can be accounted for by modifying the momentum integration measure $dp\rightarrow dp\,e^{-\ell^2 p^2}$.
This is equivalent to GUP theories for the choice $f(\vec{p}^{\, 2})=e^{\ell^2 p^2}-1$ in equation (\ref{eq:gup_comm}).
From this, we see that the Fourier transform of a momentum eigenstate is given by 
\begin{eqnarray}
\int dk\,e^{ikx} \delta(p(k)-p(k'))&=&\int dp' \left|\frac{\partial k}{\partial p'}\right|e^{ik(p')x}\frac{\delta(p-p')}{\left|\frac{\partial k}{\partial p'}\right|_{p}}\\
&=&\int dp'\, e^{-\ell^2 p'^2} e^{ik(p')x}\frac{\delta(p-p')}{e^{-\ell p^2}}=e^{ik(p)x},
\end{eqnarray}
where 
\begin{equation}
\frac{dk}{dp}=e^{-\ell^2 p^2}\Rightarrow k(p)=\frac{\sqrt{\pi}}{2\ell}\,\mathrm{erf}\left(\ell p\right)+c.
\end{equation}
with the error function $\mathrm{erf}(x)$.
The integration constant $c$ has to be zero, because $k=0$ when $p=0$.
This dispersion relation reduces to the classical case for $p\ll \ell^{-1}$ and shows a saturation behaviour for large momenta with $k(p)\stackrel{p\rightarrow\infty}{=}\frac{\sqrt{\pi}}{2\ell}$.
Generalising this to four-vectors $k^\mu=\left(\omega, \vec{k}\right)$ and $p^\mu=\left(E, \vec{p}\right)$ we have
\begin{equation}
k^\mu=\frac{\sqrt{\pi}}{2\ell}\mathrm{erf}(\ell p^\mu),
\end{equation}
and in particular
\begin{equation}
\omega(E)=\frac{\sqrt{\pi}}{2\ell}\mathrm{erf}(\ell E).
\label{eq:omega}
\end{equation}
Indeed, in \cite{hossenfelder} it is shown that the mass condition $E^2=\vec{p}^{\,2}+m^2$ remains valid if $\omega(E)$ has the same functional form as $k(p)$.
Starting from (\ref{eq:omega}), we can follow the standard derivation of the oscillation probability (Eq. (\ref{eq:osc_prob})-(\ref{eq:osc_app})).
The crucial step is given explicitly by 
\begin{eqnarray}
\omega(E_k)-\omega(E_j)&=&\frac{\sqrt{\pi}}{2\ell}\left(\mathrm{erf}\left(\ell|\vec{p}_k|\sqrt{1+\frac{m_k^2}{|\vec{p}_k|^2}}\right)-\mathrm{erf}\left(\ell|\vec{p}_j|\sqrt{1+\frac{m_j^2}{|\vec{p}_j|^2}}\right)\right)\nonumber\\
&\approx& \frac{\sqrt{\pi}}{2\ell}\left(\mathrm{erf}\left(\ell |\vec{p}_k|\right)+\frac{\ell}{\sqrt{\pi}} \frac{m_k^2}{|\vec{p}_k|}\ e^{-\left(\ell |\vec{p}_k|\right)^2}\right) \nonumber\\
&\quad& -\frac{\sqrt{\pi}}{2\ell}\left(\mathrm{erf}\left(\ell |\vec{p}_j|\right)+\frac{\ell}{\sqrt{\pi}} \frac{m_j^2}{|\vec{p}_j|}\ e^{-\left(\ell |\vec{p}_j|\right)^2}\right)\nonumber\\
&\approx& \frac{\sqrt{\pi}}{2\ell}\left(\mathrm{erf}\left(\ell E\right)+\frac{\ell}{\sqrt{\pi}} \frac{m_k^2}{E}\ e^{-\left(\ell E\right)^2}\right) \nonumber\\
&\quad& -\frac{\sqrt{\pi}}{2\ell}\left(\mathrm{erf}\left(\ell E\right)+\frac{\ell}{\sqrt{\pi}} \frac{m_j^2}{E}\ e^{-\left(\ell E\right)^2}\right)\nonumber\\
&=&\frac{\Delta m^2_{kj}}{2 E}e^{-\left(\ell E\right)^2},\label{eq:approx}
\end{eqnarray}
where we have used $|\vec{p}_k|\approx|\vec{p}_j|\approx E$ and $\Delta m^2_{kj}\equiv m^2_{k}- m^2_{j}$.
From this we find the modified transition probability
\begin{equation}
P_\ell(\nu_\alpha\rightarrow\nu_\beta)=\sum\limits_{k,j=1}^{3}U^\ast_{\alpha k}U_{\beta k}U_{\alpha j}U^\ast_{\beta j}e^{-i\frac{\Delta m^2_{kj}}{2E}\exp(-\ell^2 E^2)L}.
\label{eq:mod_prob}
\end{equation}
The two-flavour case can be obtained starting from equation (\ref{eq:2f_matrix}).
%\begin{equation}U=\begin{pmatrix}\cos\theta&\sin\theta\\-\sin\theta&\cos\theta\end{pmatrix}.\end{equation}
%The overlap of the two flavour eigenstates is then given by
%\begin{equation}\begin{split}
%&\left\langle\nu_\alpha|\nu_\beta(t)\right\rangle\\
%&={\begin{pmatrix}0& 1\end{pmatrix}\begin{pmatrix}\cos\theta&-\sin\theta \\ \sin\theta&\cos\theta\end{pmatrix}}{\begin{pmatrix}e^{i\omega_1t}&0\\0&e^{i\omega_2t}\end{pmatrix}}{\begin{pmatrix}\cos\theta&\sin\theta\\-\sin\theta&\cos\theta\end{pmatrix}}{\begin{pmatrix}1\\0\end{pmatrix}}\\
%&=\sin\theta\cos\theta\left(e^{i\omega_1t}-e^{i\omega_2t}\right)
%\end{split}
%\end{equation}
%\vskip-0.5cm
The modified transition probability reads
\begin{equation}
P(\nu_\alpha\rightarrow\nu_\beta)=\left|\left\langle\nu_\alpha|\nu_\beta(t)\right\rangle\right|^2=\sin^2(2\theta)\sin^2\left(t\frac{\omega_1-\omega_2}{2}\right),
\end{equation}
which on using approximation (\ref{eq:approx}) and $t=L$ gives
\begin{equation}
P(\nu_\alpha\rightarrow\nu_\beta)=\sin^2(2\theta)\sin^2\left(\frac{\Delta m^2L}{4 E}e^{-\ell^2 E^2} \right).
\label{eq:twoflavml}
\end{equation}
As an illustration, we plot standard and modified oscillations for the two-flavour case in figure (\ref{fig:comp_osci}).

\section{Analysis of the modified oscillation pattern}
\begin{figure}
\centering
\includegraphics[scale=1.2]{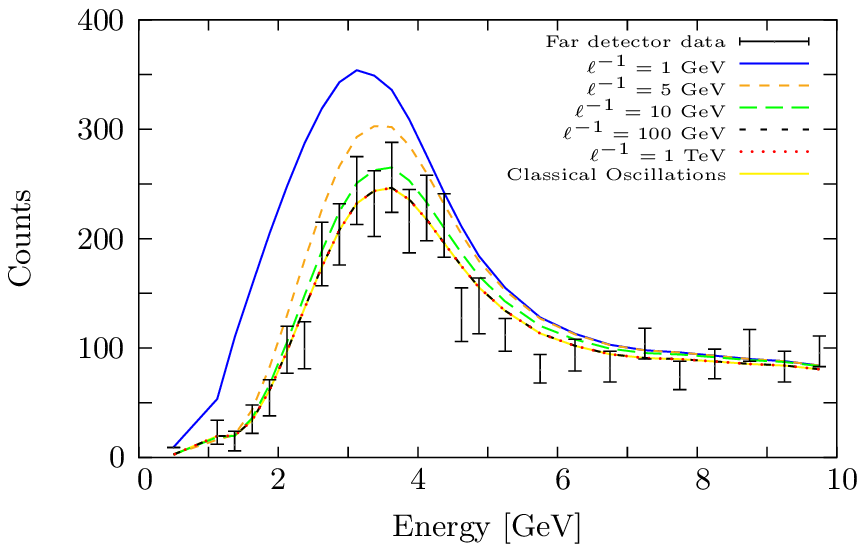}
\includegraphics[scale=0.9]{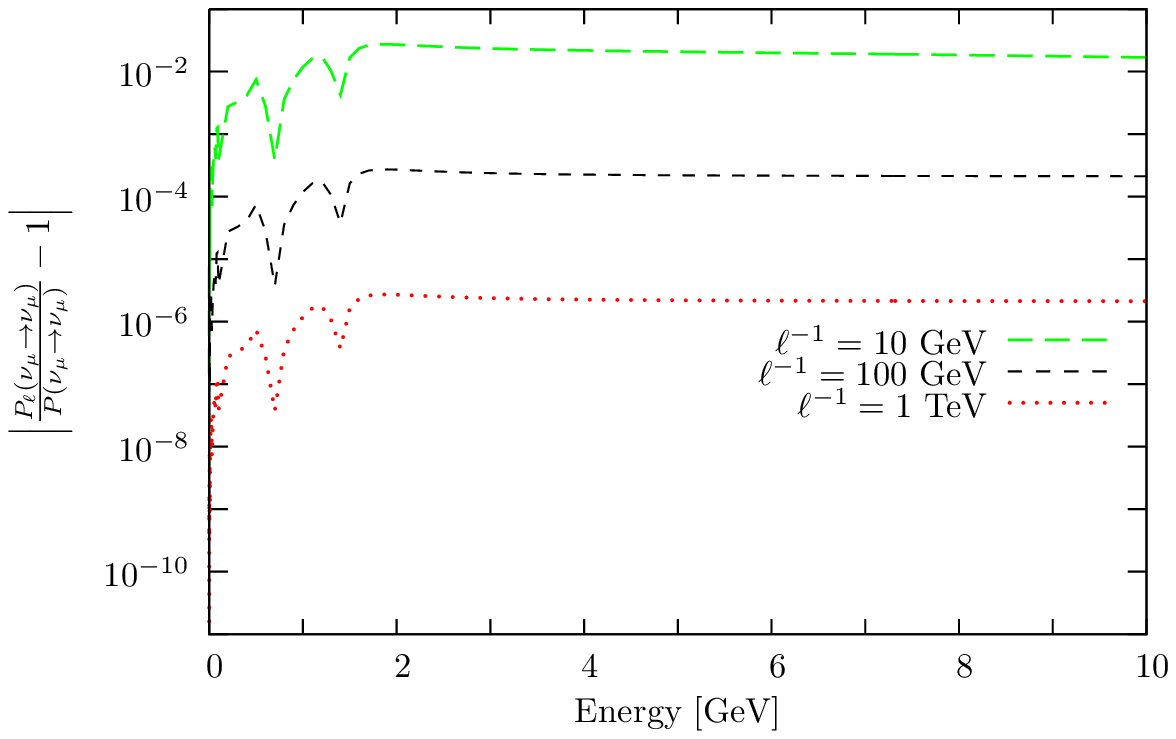}
\caption{Upper part: Expected neutrino flux for the minimal length model with different fundamental scales from $\ell^{-1}=1~\mathrm{GeV}$ to $\ell^{-1}=1~\mathrm{TeV}$ (from top to bottom) for a $\nu_\mu\rightarrow\nu_\mu$ transition with baseline $L=734~\mathrm{km}$. MINOS data taken from \cite{Adamson2011}.\\Lower part: Relative differences in the oscillation probabilities for different fundamental scales.}
\label{fig:dif_scales}
\end{figure}

Let us now analyse the modified oscillation probability Eq.(\ref{eq:mod_prob}).
Comparing our result to the literature, we find that in contrast to \cite{Lisi}, the suppression factor for the oscillation phase is not dependent on the propagation length but only on the energy.\\
The relative difference between the classical oscillations and the minimal length oscillations is shown in figure (\ref{fig:dif_scales}).
To get a very rough estimate of the value of the minimal length parameter $\ell$, we apply our modified oscillation probability Eq.(\ref{eq:mod_prob}) to available data for $\nu_\mu\rightarrow\nu_\mu$ transitions from the MINOS experiment.
As in \cite{Adamson2011}, we employ the two-flavour case Eq.(\ref{eq:twoflavml}) for the analysis using the standard oscillation prediction from \cite{Adamson2011}.
The results are shown in figure (\ref{fig:dif_scales}).
It is obvious that the differences between classical and modified behaviour are smaller than the experimental uncertainties for all realistic values of $\ell$.
From a $\chi^2$ test of the data using the modified oscillation pattern we only find a very weak bound of $\ell^{-1}\gtrsim 10~\mathrm{GeV}\,(95\%\,\mathrm{CL})$.
However, increasing the neutrino flux by a factor of $100$ might allow to push the limit into the $100~\mathrm{GeV}$ region.
Whether the size of the effect gets modified by a more sophisticated analysis is a subject of future studies.
Thus, neutrino oscillations can, in principle, provide novel information on NCG-induced minimal length pheonomena.

%END MINOS
%BEGIN LANDS
As an alternative, we look for experimental conditions which could lead to the detection of this phenomenon in the near future.
To this purpose we test the parameter space for baseline and energy range of neutrino experiments by considering a significant departure from classical oscillation probabilities, i.e.
\begin{equation}
\Delta p\equiv \left|P(\alpha\rightarrow\beta)-P_{\ell}(\alpha\rightarrow\beta)\right| > 0.1.
\end{equation}

\begin{figure}
\centering
\includegraphics[scale=1.3]{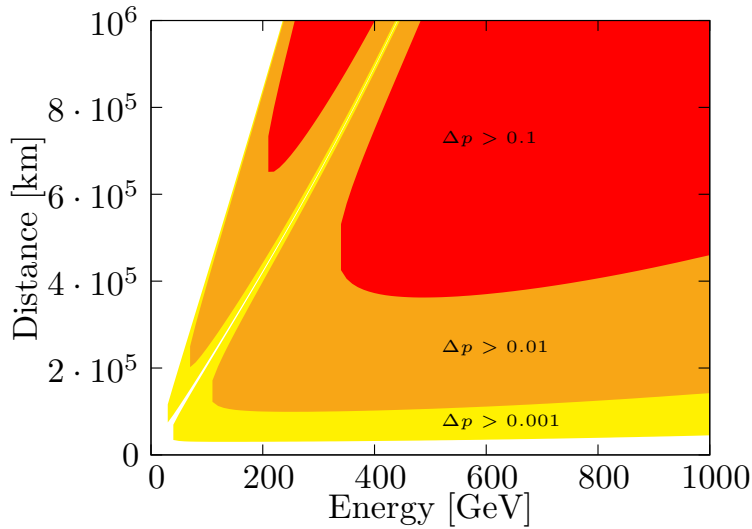}
\caption{Parameter space for three-flavour oscillations and $\ell^{-1}=1~\mathrm{TeV}$. In the indicated regions neutrino oscillations are coherent and show a significant departure from classical oscillations for different values of $\Delta p$.}
\label{fig:lands}
\end{figure}

\begin{table}[!bth]
\caption{Oscillation parameters used in the analysis of the modified oscillation behaviour. Values taken from \cite{PDG}.}
\label{tab:param}
\begin{indented}
\item[]\begin{tabular}{@{}lllll}
\br
Parameter & Value &\quad& Parameter & Value\\
\mr
$\sin^2(2\theta_{12})$&$0.861$&\quad&$\Delta m_{12}^2$&$7.59\cdot10^{-5}\,\mathrm{eV}^2$\\
$\sin^2(2\theta_{13})$&$0.15$&\quad&$\Delta m_{13}^2$&$2.43\cdot10^{-3}\,\mathrm{eV}^2$\\
$\sin^2(2\theta_{23})$&$0.92$&\quad&$\Delta m_{23}^2$&$2.43\cdot10^{-3}\,\mathrm{eV}^2$\\
$\delta_{CP}$&$0$&&\\
\br
\end{tabular}
\end{indented}
\end{table}
Fig. (\ref{fig:lands}) shows regions in which the above condition for the experimental parameters of propagation length L and neutrino energy E for a $\nu_\mu\rightarrow\nu_\mu$ transition in the three-flavour model is fulfilled, with oscillation parameters as in table (\ref{tab:param}).
Measuring neutrino oscillations in these parameter regions should lead to an observable increase or decrease of detected muon neutrinos $\nu_\mu$ in comparison to the standard oscillations.
Note that within these regions the number of neutrino counts can be larger by a factor of 10 or more for small values of the classical oscillation probability.
As we are only looking at the difference in oscillation probabilities, these results are only weakly dependent on the values of the oscillation parameters as further calculations confirm.
We see that already at distances within the solar system a significant effect could be measurable.

Looking at Eq.(\ref{eq:mod_prob}) we see that for neutrinos with energies above the fundamental scale oscillations are strongly suppressed compared to the classical case.
This is equivalent to saying that the oscillation phase in the minimal length scenario,
\begin{equation}
\frac{\phi}{2\pi}=\frac{\Delta m^2 L}{2E}\exp(-\ell^2E^2),
\end{equation}
%\textbf{[Note:Now consistent with definition above.]} 
is exponentially damped and coherence is maintained for much larger distances than in the standard oscillation framework.
This is a clear experimental signature which could be tested using astrophysical sources of high-energetic neutrinos.
Neutrinos of sufficient energy are for example created in the atmosphere by cosmic rays for which, however, the baseline is too short to provide a realistic setup.
The suppression of oscillations should also leave an imprint in the flux of high-energy cosmogenic neutrinos.
As cosmogenic neutrinos are created in flavour eigenstates $\nu_e$, $\nu_\mu$ only, neutrinos with energies greater than the fundamental scale will not oscillate, leading to a suppression of $\tau$-neutrinos in the cosmogenic flux.
This, however, is difficult to establish experimentally due to the very small flux of cosmogenic neutrinos.

On the other hand, possible point sources of high-energy neutrinos such as gamma-ray bursts or active galactic nuclei could provide the possibility to test the modified oscillation behaviour even without exact knowledge of the baseline and flavor composition of the neutrinos.
In the standard oscillation scenario, whatever the composition of the high-energy neutrino beam at the source is, the neutrinos will reach the earth in a fully mixed state (if the baseline is larger than the oscillation length which we assume is the case for the aforementioned point sources).
In contrast, in our minimal length scenario the high-energy neutrinos will reach the earth in their original flavor composition.
If neutrino telescopes such as IceCube or ANTARES indeed find a deviation from a perfect flavor mixing for high-energy neutrino spectra coming from a point source, this clearly hints at a modification of neutrino oscillations such as the one advertised in this paper and allows a test of the proposed scenario within near future.

\section{Summary}
To summarise, we have implemented an effective quantum gravity minimal length.
Modifications of the neutrino oscillation probability were calculated and show a deviation from the classical behaviour at high energies.
A rough estimation using MINOS data limits the size of the minimal length to $\ell^{-1}\gtrsim 10~\mathrm{GeV}$. A significant increase in statistics of earthbound experiments or a drastic change of the experimental baseline will allow to put stringent limits on $\ell^{-1}$.
Therefore, we have considered alternative neutrino sources, such as atmospheric, galactic, extragalactic and cosmogenic neutrinos. We showed that gamma ray bursts and active galactic nuclei could provide signals of modified oscillation probabilities due to the presence of the minimal length. Neutrinos with energies larger than the fundamental scale would propagate from the source to the Earth in their original flavour state, with a complete suppression of oscillations. This is in contrast to the standard oscillation scenario and provides a striking experimental signature that would allow to test our current understanding of quantum gravity.\\
This work is supported by the Helmholtz International Center for FAIR within the
framework of the LOEWE program (Landesoffensive zur Entwicklung Wissenschaftlich-\"{O}konomischer
Exzellenz) launched by the State of Hesse.

\end{document}